# All-Linear Multistate Magnetic Switching Induced by Electrical Current

Meiyin Yang[1], Yanru Li[1,3], Jun Luo[1,3*], Yongcheng Deng[2,3], Nan Zhang[2,3], Yan Cui[1], Peiyue Yu[1,3], Tengzhi Yang[1,3], Yu Sheng[2], Sumei Wang[1], Jing Xu[1], Chao Zhao[1,3] and Kaiyou Wang[2,3,4*]

[1]*Key Laboratory of Microelectronic Devices and Integrated Technology, Institute of Microelectronics, Chinese Academy of Sciences (IMECAS), Beijing 100029, China*
[2]*SKLSM, Institute of Semiconductors, CAS, P. O. Box 912, Beijing 100083, China*
[3]*University of Chinese Academy of Sciences (UCAS), Beijing 100049, China*
[4]*Beijing Academy of Quantum Information Sciences, Beijing 100193, China*
\* Correspondence and requests for materials should be addressed to J. L. (luojun@ime.ac.cn) and K. W. (e-mails: kywang@semi.ac.cn).

**We present an alternative mechanism to control the domain wall motion, whose directions are manipulated by the amplitude of electrical currents when modulating the ratio of D/A (constants of Dzyaloshinskii-Moriya interaction over exchange interaction). To confirm this mechanism, we observe this type of domain wall motion and demonstrate linear magnetic switching without hysteresis effect via adjusting the D/A of Ta/Pt/Co/Ta multilayer device with ion implantations. We further find field-free biased and chirally controllable multistate switching at the lateral interface of ion exposed and unexposed area, which is due to the current induced Neel wall motion and a strong exchange coupling at this interface.**

Multistate magnetic switching by current induced domain wall motion promises multiple applications in nonvolatile memories [1], logics [2] and artificial intelligence computing [3-5]. Experiments [6-10] and simulations [11-14] proved that alternating the sign of spin currents via spin-orbit torque (SOT) [15, 16], magnetic domain walls can be driven to opposite directions. Positive current drives the domain wall to one direction and negative current to the other direction, since the spin vectors of the spin currents exerted on the domain wall are determined by the sign of charge current. These domain wall motions driven by SOT exhibit hysteresis effects [9, 17-20], so that a large opposite current is required to drive the domain wall moving to a targeted direction to realize multistate switching under an in-plane field [11] or effective field [3].

To make energy efficient multistate devices, our concept is based on the modulation of competitions between exchange interaction and Dzyaloshinskii-Moriya interaction (DMI) to realize a type of linear multistate switching, where the direction of domain wall motion is driven by current amplitude. In a SOT structure (heavy metal/ferromagnets), the Dzyaloshinskii-Moriya interaction (DMI) plays a crucial role in the domain wall motion during the current induced magnetization switching, as it stabilizes the spiral domain wall, leading to increased efficiency of domain wall motions [21]. The DMI, a term written as $H_{\text{DMI}} = -D_{ij} \cdot (S_i \times S_j)$ in Hamiltonian, favors the non-collinear spin texture. In a magnetic ordered system, the DMI often competes with or is suppressed by the exchange interaction, $H_{\text{ex}} = -A_{ij} S_i \cdot S_j$, which aligns the spin in parallel or antiparallel configuration. Here, we find an alternative mechanism to tune the $D_{ij}A_{ij}^{-1}$ ratios with the assistance of nitrogen (N) ion implantations, and thus effectively manipulate the direction of domain wall motion by electric current amplitude. This type of domain wall motion is confirmed by magnetic optical Kerr effect (MOKE) and results in a linear multistate switching without hysteresis effect by current. We further construct lateral interfaces between ion exposed and unexposed areas to realize field free multistate devices. Neel walls formed at these interfaces induced the chirally controllable switching. Also, strong exchange coupling at the interfaces is also observed, which is responsible for the biased linear switching behavior .

Starting from constructions of linear multistate switching theoretically, we showed the influences of exchange interaction and DMI on the current induced magnetization switching by micromagnetic simulations. The simulation parameters are described in [31]. We simulated the $M_z$ vs. current density ($J$) pulses under an in-plane field of 200 Oe for different $D$ and $A$ constants in Figs. 1(a) and 1(b). Adjusting the $D \times A^{-1}$ values, the switching loops can be tuned from square ($D \times A^{-1} = 1.3 \times 10^8$ m$^{-1}$) to linear ($D \times A^{-1} = 3.3 \times 10^8$ m$^{-1}$). The long strip domain patterns for the larger $D \times A^{-1}$ system after each current pulse are illustrated with numbers in Fig. 1(c). The simulated results suggest that the linear $M$-$J$ loop and associated domain structure appears when the exchange stiffness is weak (large $D \times A^{-1}$ ratio). Also, at this $D \times A^{-1}$ ratio, switching loops can be controlled between square and linear shapes by changing the magnitude of in-plane external fields (Supplementary S1 in [31]). To achieve this type of linear multistate switching, we need to seek for materials with a weak exchange interaction and strong DMI.

Previous studies show that the exchange coupling was greatly weakened by ion implantations in many magnetic systems [22-24], which may be utilized to tune the ratio of $D \times A^{-1}$ to achieve a DMI dominant system. Following this guideline, we then fabricated Hall devices using thin films of Ta (3 nm)/Pt (3 nm)/Co (1 nm)/Ta (2 nm) on SiO$_2$ substrate. Subsequent N ion doses of 0 cm$^{-2}$, 3×10$^{13}$ cm$^{-2}$ and 4.5×10$^{13}$ cm$^{-2}$ were implanted into these devices under 10 keV with fixing ion beam angle of 30 degree to the film plane. For simplicity, these samples are referred as $S_1$, $S_2$ and $S_3$. To obtain their $D \times A^{-1}$ ratios, we conducted the measurements of $H_{\text{DMI}}$ (DMI fields, the saturated point of the spin-orbit torque efficiency under in-plane fields [14] in the insets of Fig. 2(a)), which allows us to access the ratios by applying the equation $H_{\text{DMI}} = D/\left(\mu_0 M_s \sqrt{A/K_{\text{eff}}}\right)$ [25]. The detailed measurements and calculations of $H_{\text{DMI}}$, $M_s$ (saturation magnetization), $K_{\text{eff}}$ (effective anisotropy constant), $D \times A^{-0.5}$ and $D \times A^{-1}$ ratios were described in supplementary S2 and S3 [31]. The $H_{\text{DMI}}$ is found to be the same for all three samples in Fig. 2(a). As seen in Fig. 2(b), both the $K_{\text{eff}}$ and $M_s$ decrease with the increase of implantation doses. The $D \times A^{-0.5}$ ratio increases from 182 J$^{0.5}$ m$^{-1.5}$ ($S_1$) to 258 J$^{0.5}$ m$^{-1.5}$ ($S_3$) in Fig. 2(c). The bottom limits of $D \times A^{-1}$ ratios were also plotted, which doubles from 4.2×10$^7$ m$^{-1}$ to 8.4×10$^7$ m$^{-1}$ as the dose increases from 3×10$^{13}$ cm$^{-2}$ to 4.5×10$^{13}$ cm$^{-2}$. Therefore, for samples with ion implantations, the $A$ values are severely lowered and $D \times A^{-1}$ increases remarkably.

The simulations predicted that weak exchange stiffness and large $D \times A^{-1}$ ratio would induce the linear magnetic switching and the experiments have shown that ion implantations could enable specific materials with such parameters. We then examined their electric transport properties by measuring Hall devices as illustrated in Fig. 3(a). Both $S_1$ and $S_3$ samples have perpendicular anisotropy in Fig. 3(b). The Hall resistance ($R_H$) of $S_1$ can be switched sharply by current pulses (500 ms) under an external in-plane $H_x$ magnetic field (Fig. 3(c)), which is similar to previous results [26-28]. With ion implantations, multistate $R_H$ exhibits perfectly linear behavior controlled by the current pulses and no hysteresis loop is displayed in Fig. 3(d). The linear reduction of $R_H$ indicates that the domain wall motion can be reversed by adjusting current amplitudes. To investigate the associated domain wall motion, we probed the domain wall structures under current pulses using MOKE. Fig. 3(e) shows the domain images of sample $S_3$ ($4.5\times10^{13}$ cm$^{-2}$) when the current was swept from 4.5 mA to −4.5 mA under an in-plane field of −220 Oe. Lowering the current pulses, reversed domains grow into strips patterns, indicating that the system has strong DMI [29], so that the domain walls move faster along $H_x$ ($x$ axis) direction than that along $y$ axis by the SOT. These domain wall patterns resemble simulated domains in Fig. 1(c).

In addition, the linear magnetic switching of both $S_2$ and $S_3$ samples takes place under smaller in-plane fields (Supplementary S5 in [31]). At larger in-plane fields, the linear switching can be modulated into hysteresis loops in Fig. S5(a) and Fig. S6(a) [31]. These experimental observations are perfectly consistent with the simulations in Fig. 1 and Fig. S1 [31], further proving the mechanism of linear magnetic switching.

Although the linear current induced multistate switching can be realized with the assistance of ion implantations, the external fields are still necessary. To get rid of this field, we broke the spatial symmetry of Hall devices using a photoresist mask that covering half of the Hall channel as shown in Fig. 4(a). Then it was exposed to N ion implantations with a dose of $4.5\times10^{13}$ cm$^{-2}$. A two-step switching in Fig. 4(b) indicates that the $H_C$ value in the area with ion implantations is smaller with easy axis pointing along $z$ axis. To simplify the descriptions, the moments of the N ions exposed area and the unexposed area are symbolized to be $m_N$ and $m$. After positive or negative perpendicular preset fields, we swept the smaller $H_z$ field to assess the magnetization switching of the exposed area in Fig. 4(c). The loops shift toward −$H_z$ and $H_z$ for $m\uparrow$ and $m\downarrow$ preset configurations. It means that it is easy for an external field to drive $m_N$ to the direction of $m$, indicating that $m_N$ and $m$ are strongly lateral ferromagnetic coupled at the interface between exposed and unexposed areas. Such a lateral coupling with a DMI chirality were also revealed recently in a nano size device [30]. For our devices, the lateral exchange coupling is able to affect devices up to a micro size.

Field-free deterministic switching by currents were clearly observed in Fig. 4(d) when injecting current of $I_1$ along $x$ axis, where the symmetry-breaking is perpendicular to the current. The larger switching current corresponds to the switching of *m*, and the smaller one is associated with the switching of $m_N$. We also measured the minor switching loop of $m_N$ by fixing the *m*, which was preset by a large $H_z$ field. Deterministic switching was then observed in Fig. 4(e). Changing the direction of *m* does not alter the chirality of the switching. We flipped the device clockwise by 90 degree and performed the same measurements, *i.e.* injecting current $I_2$ along –*y* axis and measured the Hall voltages in *x* axis. We did not observe the deterministic switching behavior for larger current sweeping as depicted in Fig. 4(d), since the symmetry-breaking is parallel to the current direction. Interestingly, the minor loops behave completely different from that of injecting $I_1$ current, where the deterministic switching depends on the moment of *m*. The *m*↑ (*m*↓) results in the clockwise (anti-clockwise) multi-state switching loops in Fig. 4(f). This chirality phenomenon can be explained if the Neel walls were formed during the switching as illustrated in Fig. 4(g). In this scenario, the magnetization in the wall ($m_d$) is along *x* axis, and the effective field $m_d \times \sigma$ exerted on the wall is along *z* axis by the spin orientation $\sigma$ (along *y* axis). Thus, for a fixed chirality of domain walls, the spin-orbit torque is able to drive the domain walls deterministically. Reversing the *m* of unexposed area, the $m_d$ in the wall also switches its sign, giving rise to opposite switching behaviors deterministically.

We also noticed that all switching loops in Figs. 4(e) and 4(f) exhibit an obvious shift with the combination of hysteresis and linear multistate switching. For instance, in Fig. 4(e), the loop is shifted to the direction of positive current for *m*↑. The dash line plotted in Fig. 4(e) divides an obvious hysteresis and linear loop obviously. The strong exchange coupling at the interfaces between exposed and unexposed areas is responsible for these two phenomena. The switching not only happens in the exposed area, but also occurs at the unexposed area due to the exchange coupling effect. Owing to the coupling, the moment near the lateral interface in the unexposed area can be affected by the $m_N$ and switched by smaller current. Therefore, the switching at unexposed areas exhibits a hysteresis below the dash line. When the magnetic switching is at the exposed area, this typical linear behavior takes place (above the dash line). The magnetization in the unexposed area cannot be switched when the current is reduced and the $R_H$ remains the same (section 1). As the increase of the positive current, the $R_H$ switched abruptly (section 2). When the domain wall moves into the exposed area, this typical linear switching loop takes place (section 3). The switching of devices in section 4 is slower than that in section 2, even the switching happens in the same unexposed area, which is due to the exchange coupling between exposed and unexposed areas. When the $m$ is up, the $m_N$ is inclined to stay at the up states and therefore the magnetic switching from down to up (section 4) is easier than that from up to down (section 2).

To further verify that Neel Walls are responsible for the field-free multistate switching, systematic investigations on the Hall devices covered by photoresist with its edge angle about ±3 degree to the Hall channel in Figs. 5(a) and 5(b) were performed. As expected, the major switching loops in Figs. 5(c) and 5(d) reveal similar switching chirality for the unexposed area regardless of the angles. However, the switching behaviors in the exposed area are totally different. The switching loop of $m_N$ is clockwise for $m\uparrow$ and anti-clockwise for $m\downarrow$ when the degree is −3 degree in Fig. 5(e). We measured the domain structures of this device in supplementary S6 in [31], proving that the domain wall is indeed formed at the interface of exposed and unexposed areas and it is driven by current pulses. Interestingly, the angle of 3 degree changes the chirality loop to anti-clockwise for $m\uparrow$ and clockwise for $m\downarrow$ in Fig. 5(f). This is because the projection of $m_d$ along $x$ axis is reversed for these two samples (illustrated in Figs. 5(a) and 5(b)), resulting in different chirality deterministic switching. We further flipped these two devices by 90 degree and also observed the chirality dependence switching on the $m$ in unexposed area (supplementary S7 in [31]), because the $m_d$ also has a projection along $x$ axis in this case.

We demonstrated the controllable excitatory/inhibitory synaptic functions by the same current values using the device in Fig. 5(a). Due to the biased chirality loops, the device can be operated at ultralow current densities, $8.3\times10^5$ A cm$^{-2}$ (0.5 mA). As shown in Fig. 5(g), the black (red) curves indicate the inhibitory (excitatory) function characterized by decreased (increased) $R_H$ values after each pulse of 0.5 mA, initialized by a current of 12 mA (−12 mA) and then −2 mA. Our demonstration could pave a way to the programmable and energy efficient artificial intelligence applications with ultralow energy consumption.

This work was supported by National Key R&D Program of China (Grant No. 2017YFB0405700) and NSFC Grant (No. 11604325), Strategic Priority Research Program of Chinese Academy of Sciences, Grant Nos. XDA18000000, XDPB12 and XDB28000000,

Chinese Academy of Sciences, Grant No.QYZDY-SSW-JSC020, and the Youth Innovation Promotion Association of CAS under Grant No. 2015097.

M. Yang and Y. Li contributed equally to this work.

**Figures and figure captions**

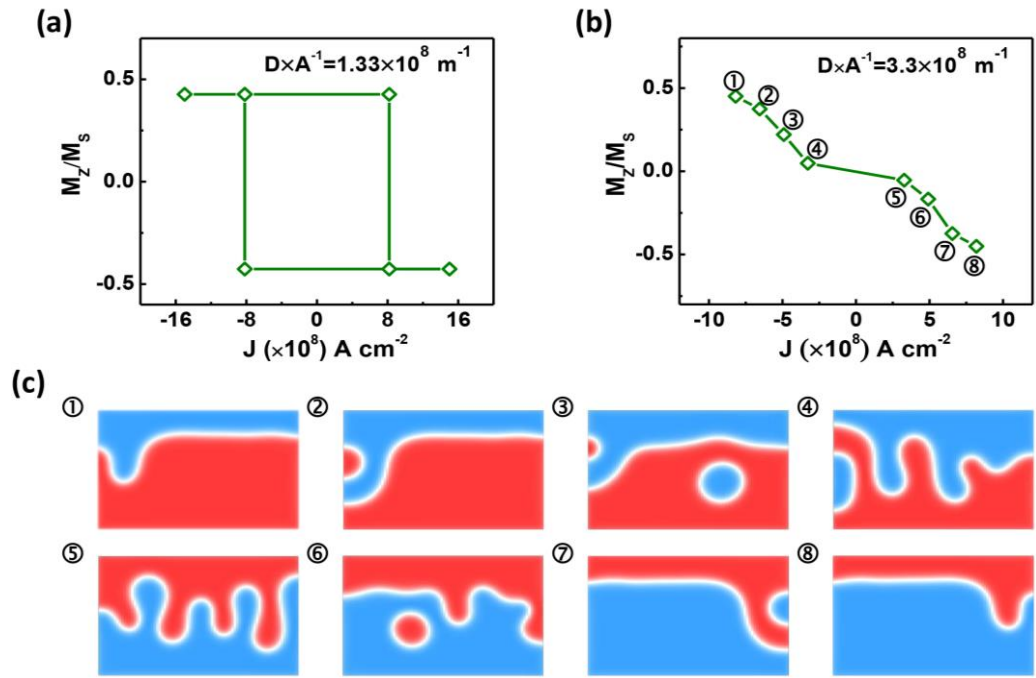

FIG. 1 (color on line). (a) The magnetic switching loop after charge current pulses by micromagnetic simulation using the $D{\times}A^{-1} = 1.3{\times}10^8$ m$^{-1}$ ($A=1.5{\times}10^{-11}$ J m$^{-1}$ and $D=2{\times}10^{-3}$ J m$^{-2}$). (b) The linear switching of $M_z$ vs. $J$ loop simulated using $D{\times}A^{-1} = 3.3{\times}10^8$ m$^{-1}$ (the $A$ is set to $0.6{\times}10^{-11}$ J m$^{-1}$ and $D$ value is retained). (c) The domain wall patterns after current pulses indicated in (b) with $D{\times}A^{-1} = 3.3{\times}10^8$ m$^{-1}$.

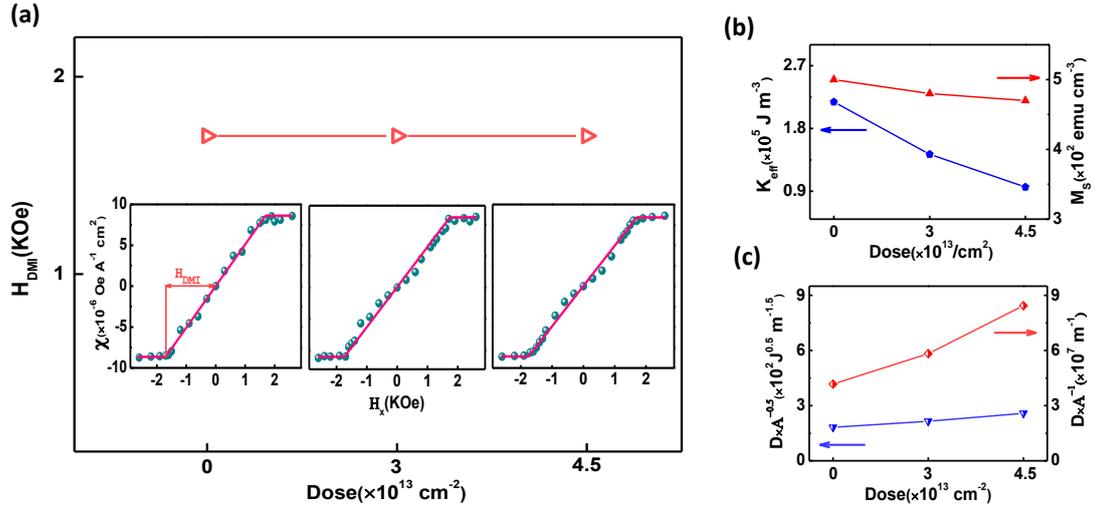

FIG. 2 (color on line). (a) The measured $H_{DMI}$ field dependence on implantation dose. The insets are the spin torque efficiencies that were measured under different in-plane fields. The saturation points are the $H_{DMI}$. (b) The calculated $K_{eff}$ and measured $M_s$ at different implantation doses. (c) The calculated $D \times A^{-0.5}$ and the bottom limits $D \times A^{-1}$ ratios at different implantation doses. The $D \times A^{-1}$ ratios of samples with ion implantation are not smaller than this value. More detailed information can be found in supplementary S3 [31].

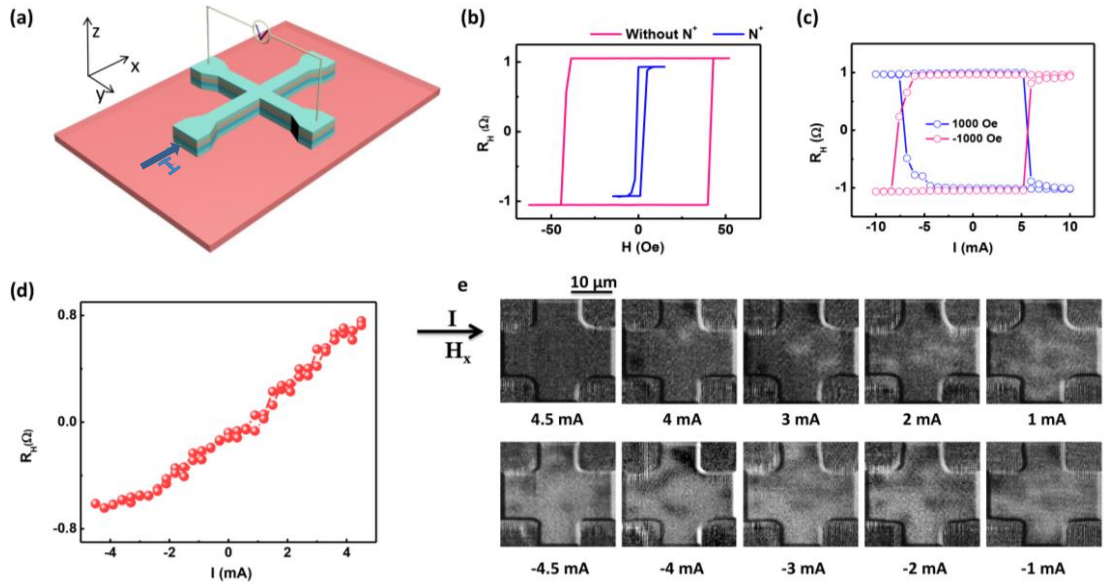

FIG. 3 (color on line). (a) Ta/Pt/Co/Ta Hall device schematics, coordinate systems and the electrical measurement set-up. (b) Hysteresis loops under perpendicular magnetic field of the devices with implantation doses of $4.5\times10^{13}$ cm$^{-2}$ and 0 cm$^{-2}$ respectively. (c) $R_H$ recorded after current pulses of 500 ms under in-plane external fields without ion implantations. (d) Pulse current induced linear multistate switching with ion implantations while applying an in-plane field of −220 Oe. (e) The domain images of the devices with ion implantation after sweeping each current pulse.

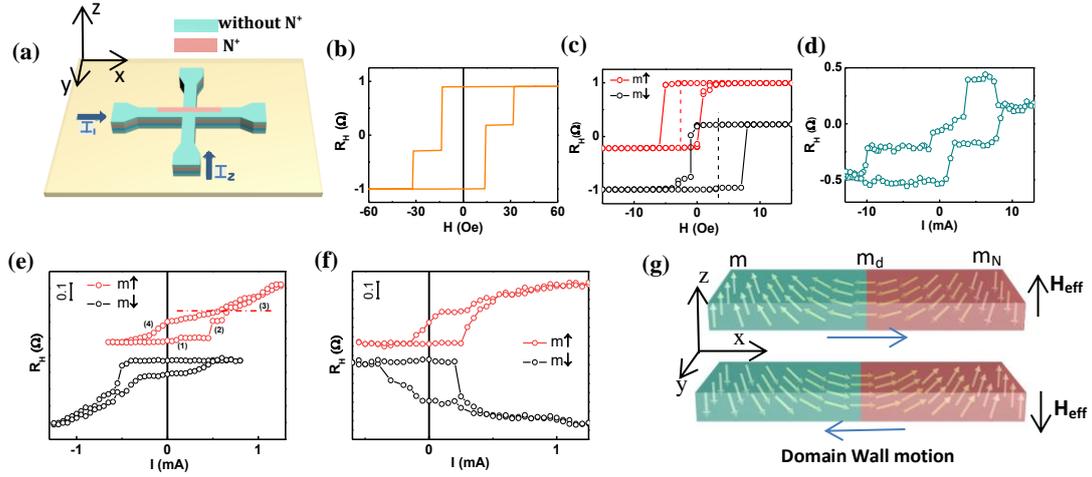

FIG. 4 (color on line). (a) The schematics of devices with half area ion exposed. The pink (blue) indicates the exposed (unexposed) areas. (b) The $R_H$ of the devices in (a) measured under perpendicular magnetic fields. (c) The minor $R_H$ loop when the moment of the unexposed area is polarized to be up and down under perpendicular magnetic fields. (d) Current induced magnetization switching without magnetic fields by injecting current pulse of $I_1$ along $x$ axis. (e) Multistate switching loops by current pulses when the moment in the unexposed area is fixed at up and down. (f) Multistate switching loops after injecting current pulse of $I_2$ to the Hall device. (g) Domain wall structure schematics when the unexposed region's moment is up or down during the current switching. According to the switching chirality in (e)-(f), the domain wall should be right-handed. Preset $m$ to ↑, the $m_d$ in the domain wall is along $-x$ direction, leading to an effective $+H_z$ for the positive current. This situation is reversed when the $m$ is ↓, resulting in the $-H_z$ effective field.

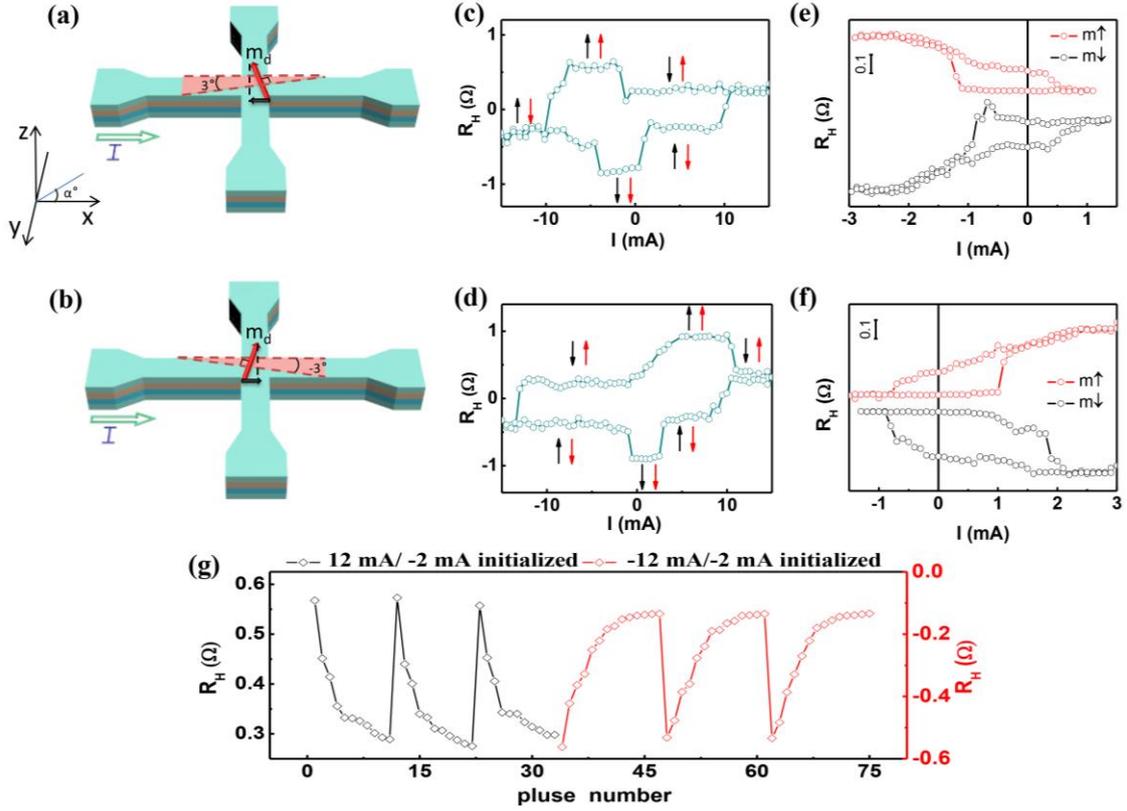

FIG. 5 (color on line). (a), (b) The schematics of devices with photoresist covered about 3 (a) and –3 (b) degree compared to the Hall channel. The angle is defined between the lateral interfaces and the *x* axis. The angle in these schematics are more than 3 degree to make the wedge clearer. The pink area is implanted with N ions. (c), (d) The field-free $R_H$ switching loops by current pulses corresponding to the devices in (a) and (b). (e) The minor switching loop by injecting small current pulses for the device in (a) for both *m↑* and *m↓*. (f) The minor switching loop of the device in (b) for both *m↑* and *m↓*. (g) excitatory/inhibitory synaptic functions with the same current pulse of 0.5 mA using the device in (a).

Supplementary Information

# Linear Multistate Magnetic Switching Induced by Electrical Current

Meiyin Yang[1], Yanru Li[1,3], Jun Luo[1,3*], Yongcheng Deng[2,3], Nan Zhang[2,3], Yan Cui[1], Peiyue Yu[1,3], Tengzhi Yang[1,3], Yu Sheng[2], Sumei Wang[1], Jing Xu[1], Chao Zhao[1,3] and Kaiyou Wang[2,3,4*]

[1] *Key Laboratory of Microelectronic Devices and Integrated Technology, Institute of Microelectronics, Chinese Academy of Sciences (IMECAS), Beijing 100029, China*

[2] *SKLSM, Institute of Semiconductors, CAS, P. O. Box 912, Beijing 100083, China*

[3] *University of Chinese Academy of Sciences (UCAS), Beijing 100049, China*

[4] *Beijing Academy of Quantum Information Sciences, Beijing 100193, China*

* Correspondence and requests for materials should be addressed to J. L. (luojun@ime.ac.cn) and K. W. (e-mails: kywang@semi.ac.cn).

**Supplementary Information**

S6. Current induced domain wall motion of half ion implanted devices.

S7. The chirality switching loops of samples in Fig. 5 which were rotated clockwise by 90 degree.

## S1. Micromagnetic simulations of the large $D \times A^{-1}$ ratio under different in-plane magnetic fields.

Micromagnetic simulations were performed using the Object Oriented MicroMagnetic Framework (OOMMF) software. We used a cuboid with 80 nm long, 50 nm wide and 1 nm thick, which was divided into meshes to a size of 2 nm×2 nm×1 nm. The $M_s$, $K_u$, spin Hall angle and damping were set to be 1000 emu cm$^{-3}$, 1×10$^6$ J m$^{-3}$, 0.3 and 0.5, respectively. The easy axis is along $z$ axis and the spin vector is pointed along $y$ axis. The endurance of the current pulse is 20 ps, and the magnetization is then relaxed to its equilibrium state after this pulse.

We find that lowering the exchange stiffness while maintaining the DMI constant (a large $D \times A^{-1}$ ratio) can lead to the linear $M_z$ dependence on the current pulses. How the current induced magnetization switching under different external in-plane fields were also calculated. Fig. S1 shows the $M_z$-$J$ loops using the parameters of $A$=6×10$^{-12}$ J m$^{-1}$ and $D$=2×10$^{-3}$ J m$^{-2}$ ($D \times A^{-1}$ =3.3×10$^8$ m$^{-1}$). When the in-plane magnetic field is set to 1000 Oe, the $M_z$ vs. $J$ loop is hysteresis in Fig. S1(a). Reducing the in-plane field to 200 Oe, the $M_z$ changes linearly after the sweeping of current pulse in Fig. S1(b). Thus, the in-plane external field can tune the shape of current switching loops effectively.

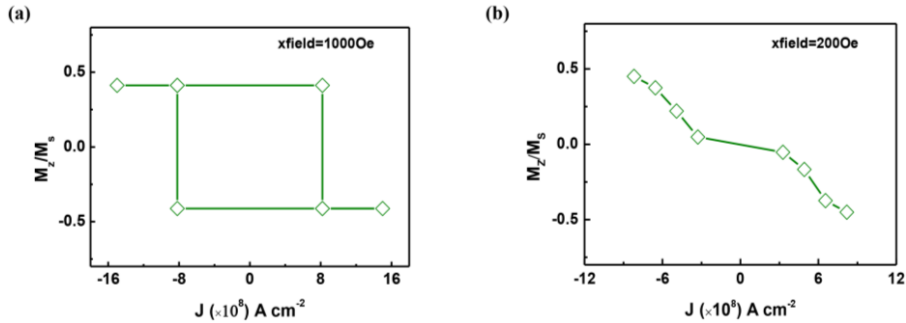

Figure S1: (a), (b), The $M_z$ vs. $J$ loop under in-plane fields of 1000 Oe (a) and 200 Oe (b) respectively.

## S2. The measurements of spin-orbit torque efficiencies at different ion implantation doses.

The magnetic switching in heavy metal/ferromagnet system can be explained by SOT+DMI scenario. Under $H_x$ fields, the spin current exerted on the magnetization induces a perpendicular effective field $H_{eff}$, which leads to a shift of $R_H$-$H$ loop as shown in Fig. S2(a). This $H_{eff}$ linearly depends on the charge current injecting in the Hall channel (Fig. S2(b)). The slope of this line is the spin-torque efficiency in the insets of Fig. 2(a).

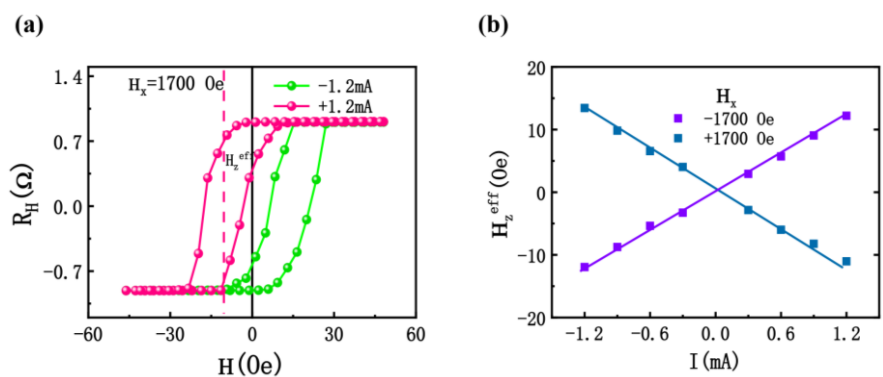

Figure S2: (a) The shift of $R_H$-$H$ loop under an in-plane magnetic field of 1700 Oe. (b) The effective field dependence on the current under in-plane magnetic fields.

The DMI stabilizes the Neel domain wall, which increases the efficiency of domain wall motion during the spin-orbit torque switching. Under the external $H_x$ field, the magnetization of the domain wall tends to align with this field, leading to a perpendicular effective field by SOT. Therefore, the $H_x$ can improve the spin-orbit torque efficiency. When the magnetization in the domain wall is fully along with the $H_x$, the spin-orbit torque efficiency does not increase any more. Fig. S2(b) is the saturated spin-orbit torque efficiency under an external field of 1700 Oe.

## S3. Magnetic hysteresis loops and $D \times A^{-1}$ ratios of Ta/Pt/Co/Ta thin films with different ion implantation doses.

To obtain the $M_s$ and $H_k$ parameters, we measured $M$-$H$ loops of Ta/Pt/Co/Ta thin films. Fig. S3(a) shows the loops of magnetization as a function of the perpendicular magnetic field. The coercivity of the thin films is lowered significantly by ion implantations, while the $M_s$ only drops a little. Figs. S3(b)-(d) illustrates the sweeps of *M-H* loops under in-plane magnetic fields with different ion implantation doses. The arrow indicates the $H_k$ values we subtract. We can see that the $H_k$ decreases with the increase of ion implantation doses.

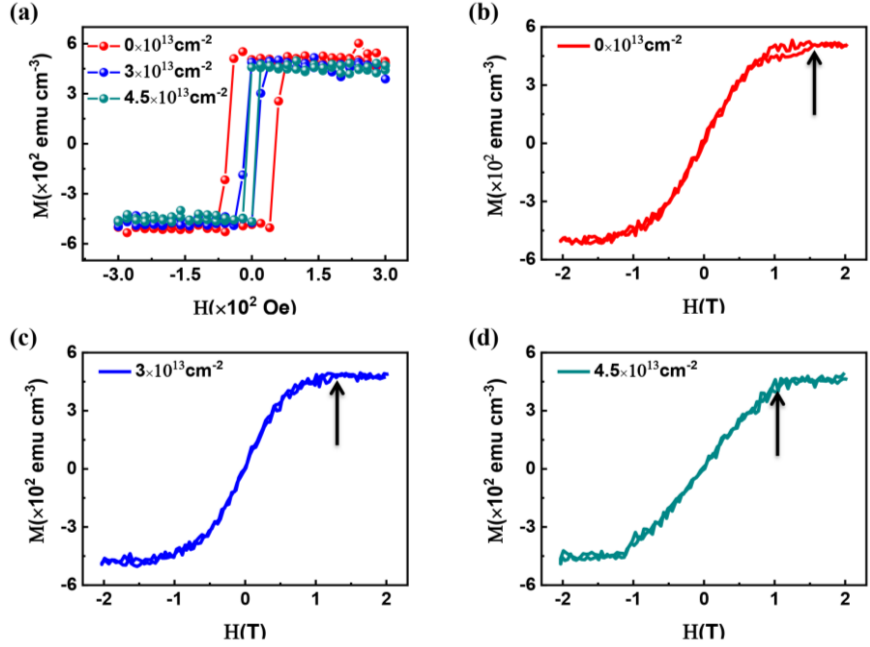

Figure S3: Hysteresis loops of Ta/Pt/Co/Ta films. (a) *M-H* loops under perpendicular magnetic fields. (b)-(d) *M-H* loops under in-plane magnetic fields of films without ion implantations (b) with implantation doses of $3\times10^{13}$ cm$^{-2}$ (c) and $4.5\times10^{13}$ cm$^{-2}$ (d).

The relationship of $H_{DMI}$ with $D$ and $A$ are given by $H_{DMI} = D/(\mu_0 M_s \sqrt{A/K_{eff}})$ (*ref* [1]), where $M_s$ is the saturation magnetization, $\mu_0$ is the vacuum permeability and $K_{eff} = K_u - \mu_0 M_s^2/2$, where $K_u$ is the uniaxial anisotropy constant. The $M_s$ was subtracted from the perpendicular hysteresis loops as shown in Fig. S3(a). The $K_u$ value was calculated using the formula $K_u = H_k M_s/2$, where $H_k$ is the magnetic anisotropy field, which was obtained from the saturation field of the in-plane hysteresis loops in Figs. S3(b)-(d). As seen in Fig. 2(b), the $K_{eff}$ decreases tremendously with the increase of implantation doses. Inputting the $K_{eff}$ and $M_s$ into the formula of $H_{DMI}$, the ratios of $D \times A^{-0.5}$ can be calculated as was plotted in Fig. 2(c). As the increase of ion implantation doses, the ratios of $D \times A^{-0.5}$ increase from 182 J$^{0.5}$m$^{-1.5}$ (S$_1$) to 258 J$^{0.5}$m$^{-1.5}$ (S$_3$). For the sample S$_1$ (without implantation), the $D$ is calculated to be $7.93\times10^{-4}$ J m$^{-2}$, which is in the range of previous reports in the system of Pt/Co thin films [2], using the exchange stiffness of Co, $A=1.9\times10^{-11}$ J m$^{-1}$ (*ref* [3]). To get the ratio of $D\times A^{-1}$, we make an assumption that the ion implantations cannot lead to increased $A$ and $D$. It is reasonable in that implanted N ions serve as impurities to thin films, which unavoidably damage the crystal structure. $A$ and $D$ are not likely to increase in such a disordered system. As a result, either $A$ or $D$ will remain unaltered or decrease after ion implantations. In this case, when $D\times A^{-0.5}$ is increasing, it means that the drop of $A$ prevails over that of $D$, as $A$ is under the square root. We then calculated the bottom limits of $D\times A^{-1}$ value for each ion implantation dose by fixing the $D$ value ($7.93\times10^{-4}$ J m$^{-2}$) in Fig. 2(c). The $D\times A^{-1}$ doubles from $4.2\times10^7$ m$^{-1}$ to $8.4\times10^7$ m$^{-1}$ as the dose increases from $3\times10^{13}$ cm$^{-2}$ to $4.5\times10^{13}$ cm$^{-2}$. If $D$ decreases with the increase of implantation doses, the $D\times A^{-1}$ should be larger than that with fixed $D$ value. Therefore, the $D\times A^{-1}$ ratios of the samples with ion implantation can not be smaller than these bottom limits. The $D$ in this system does not likely reduce significantly, as the transmission electron microscopy (TEM) images of $S_1$ and $S_3$ in supplementary S4 show similar Pt/Co/Ta interfaces.

**S4. Transmission Electron Microscopy (TEM) images of Ta/Pt/Co/Ta thin films.**

The cross-sectional TEM images and Energy Dispersive X-Ray Spectroscopy (EDS) images of Ta/Pt/Co/Ta thin films with or without ion implantations are shown in Fig. S4. We do not receive the N signal by EDS, due to the small N percentage in thin films. No obvious damage at Pt/Co/Ta interfaces with N implantation of $4.5\times10^{13}$ cm$^{-2}$ is observed in Fig. S4(b).

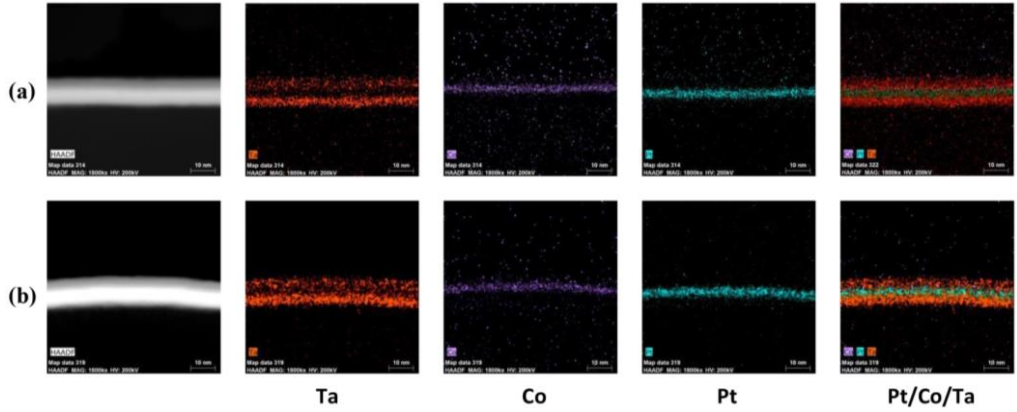

Figure S4: (a), (b) The TEM images of Ta/Pt/Co/Ta without implantation (a) and with implantation to a dose of $4.5\times10^{13}$ cm$^{-2}$ (b).

## S5. Current induced magnetic switching of samples with different ion implantation doses.

Simulations predict that the *M-J* loops could be tuned by in-plane fields between square and linear shape at a larger $D\times A^{-1}$ ratio. We then performed the current switching measurement with different in-plane fields using the samples with ion implantations. Fig. S5 and Fig. S6 show the switching loops of devices with implantation doses of $3\times10^{13}$ cm$^{-2}$ and $4.5\times10^{13}$ cm$^{-2}$ under in-plane fields respectively. For devices with a smaller dose in Fig. S5, the switching loop is very square under the fields of ±287 Oe and ±212 Oe. Lowering the field from 112 Oe to 55 Oe, the loop becomes linear after large current pulses and square at small current pulses. For devices with a larger dose in Fig. S6, a strong field of about –1130 Oe is required to realize a square switching loop. Under the field of –340 Oe, the hysteresis loop almost disappears. Further lowering the in-plane fields to –220 Oe, a perfect linear switching is achieved. We can see that as the increase of implantation doses, the linear switching is easy to be realized with smaller in-plane fields.

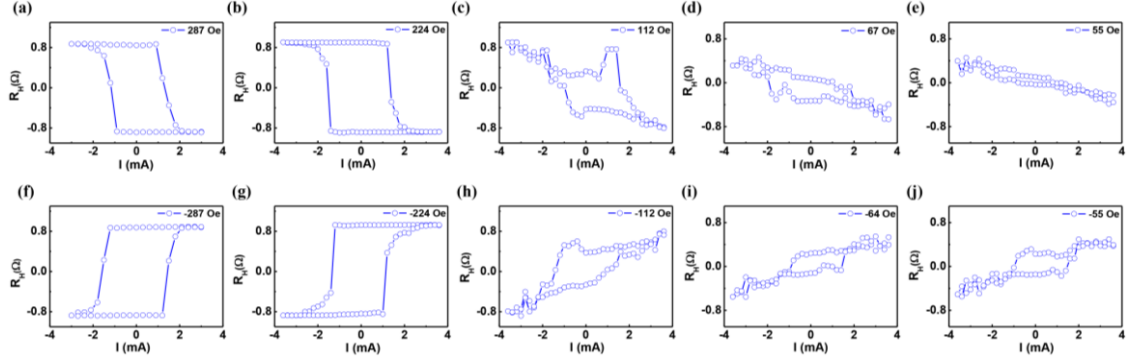

Figure S5: (a)-(e) current induced switching under positive in-plane fields from 287 Oe to 55 Oe. (f)-(j) negative in-plane fields from –287 Oe to –55 Oe of the device with an implantation dose of $3\times10^{13}$ cm$^{-2}$.

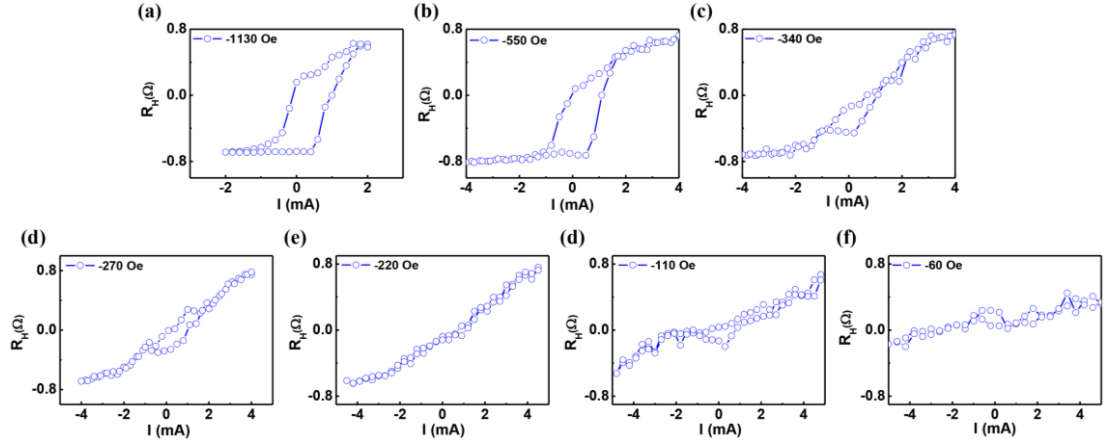

Figure S6: (a)-(f) current induced switching under negative in-plane fields from –1130 Oe to –60 Oe for devices with an implantation dose of $4.5\times10^{13}$ cm$^{-2}$.

**S6. Current induced domain wall motions of half ion implanted devices**

To explain why different chiralities of switching loops rely on the moment of unexposed area, we measure the domain wall patterns of the devices in Fig. 5(a). We first polarized the devices by a large negative perpendicular magnetic field, and then applied current pulses of –2 mA, –1.5 mA, –0.5 mA, 0.5 mA and 1 mA. After each current pulse, domain structures were scanned (Figs. S7(a)-(e)). At –2 mA, the magnetization of most areas is in the -z axis, except the bottom black strip whose magnetization is along +z axis. Further lowering the current, the black domain grows along y axis, with the domain wall nearly along x axis. The domain wall is parallel to the ion exposed and unexposed interfaces. The field-free multistate memory by SOT for this device is due to the domain wall motion driven by the currents. We find that the lateral interface of ion exposed and unexposed area prefers to form domain walls due to different magnetic properties. The domain wall at the interface is Neel wall. The magnetization in the wall has projection along *x* axis. By the spin current generated by SOT, this kind of domain wall can move along *y* axis by injecting current along *x* axis.

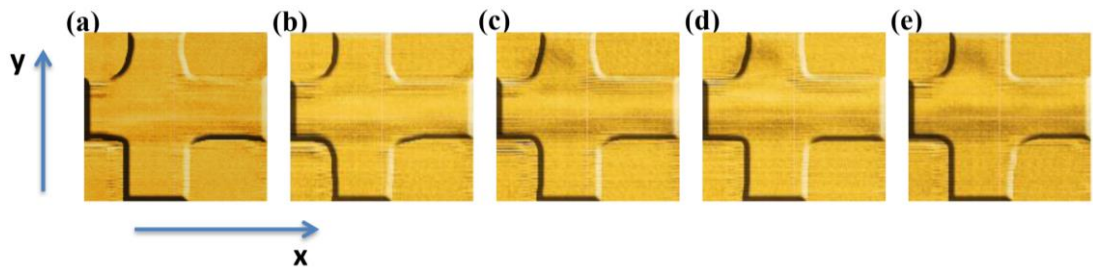

Figure S7: Domain wall motion of the device with partially ion implantation of $4.5\times10^{13}$ cm$^{-2}$ after current pulses of (a) –2 mA; (b) –1.5 mA; (c) –0.5 mA; (d) 0.5 mA; (e) 1 mA.

## S7. The chirality switching loops of devices in Fig. 5 which are rotated clockwise by 90 degree.

The domain wall at the interface of exposed and unexposed areas is regarded as Neel wall in order to explain the chirality switching loops. To further explore this in detail, we flipped the devices in Fig. 5 clockwise by 90 degree and measured its switching loops. The angle of the ion exposed and unexposed interface with the Hall channel becomes to be 93 degree of 87 degree after flipping the devices by 90 degree as illustrated in Figs. S8(a) and S8(d). In this case, because the symmetry-breaking is nearly parallel to the current direction, we do not observe the deterministic switching of the unexposed area. After we magnetized the exposed area to up or down direction, the deterministic switching of the exposed area with opposite chirality can be observed in Figs. S8(b) and S8(c). We also notice that the switching chiralities are different for the devices with 93 degree and 3 degree (Fig. 5(e)) when the magnetization in unexposed area is fixed to up and down. This is because the projection of magnetization in the domain wall along the current axis is opposite for these two angles. For the devices with an angle of 87 degree and –3 degree (Fig. 5(f)), the situation is similar. They show different chirality switching loops with each other.

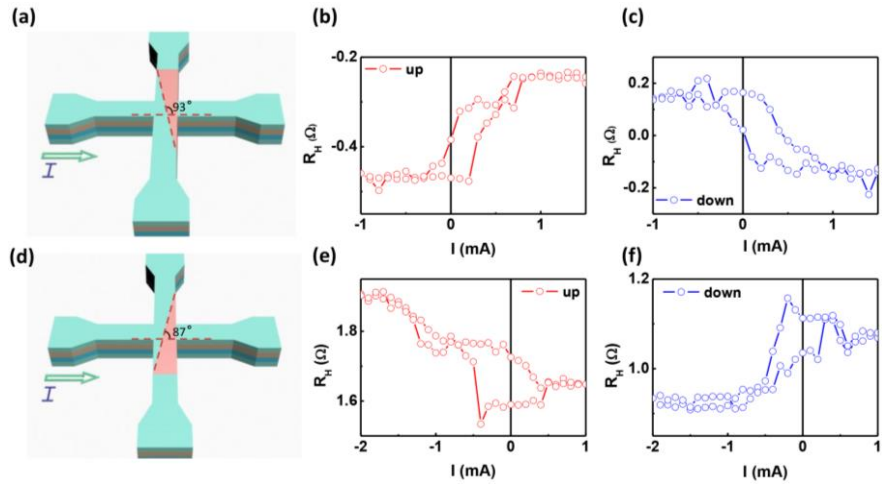

Figure S8: Chirality switching loops when flipping the devices by 90 degree clockwise in Fig. 5(a) and Fig. 5(b). (a) The device schematics after rotating the device clockwise by 90 degree in Fig. 5(a). The angle of the interfaces with the Hall channel becomes to be 93 degree. (b), (c) Current induced switching loops for the 93 degree sample when the unexposed region is up (b) and down (c). (d) The device schematics after rotating the device clockwise by in Fig. 5(b). The angle of the interfaces with the Hall channel becomes to be 87 degree. (e), (f) current induced switching loops for the 87 degree sample when the unexposed region is up (b) and down (c).